\documentclass[aps,prl,twocolumn,superscriptaddress]{revtex4-2}

\usepackage{gensymb}  
\usepackage{graphicx}  
\usepackage{hyperref}

\hypersetup{colorlinks=true,urlcolor= blue,citecolor=blue,linkcolor= blue}

\usepackage{enumitem}
\setlist{noitemsep,leftmargin=*,topsep=0pt,parsep=0pt}
\usepackage{chemmacros}

\begin{document}


\title{Lattice-Driven Chiral Charge Density Wave State in 1T-TaS$_{2}$}


\author{Manoj Singh}
\author{Boning Yu}
\author{James Huber}
\author{Bishnu Sharma}
\author{Ghilles Ainouche}
\author{Ling Fu}
\affiliation{Department of Physics, Clark University, Worcester, Massachusetts 01610, USA}

\author{Jasper van Wezel}
\affiliation{Institute for Theoretical Physics Amsterdam, University of Amsterdam, Science Park 904,1098 XH Amsterdam, The Netherlands}

\author{Michael C. Boyer}
\email{To whom correspondence should be addressed. mboyer@clarku.edu}
\affiliation{Department of Physics, Clark University, Worcester, Massachusetts 01610, USA}

\date{\today}

\begin{abstract}
We use scanning tunneling microscopy to study the domain structure of the nearly-commensurate charge density wave (NC-CDW) state of 1T-TaS$_2$. In our sub-angstrom characterization of the state, we find a continual evolution of the CDW lattice from domain wall to domain center, instead of a fixed CDW arrangement within a domain. Further, we uncover an intradomain chirality characterizing the NC-CDW state. Unlike the orbital-driven chirality previously observed in related transition metal dichalcogenides, the chiral nature of the NC-CDW state in 1T-TaS$_2$ appears driven by a strong coupling of the NC-CDW state to the lattice.
\end{abstract}

\maketitle

\section{}

The complexity of the physics driven by electron-electron and electron-phonon interactions makes 1T-TaS$_2$ a fertile playground to study complex orders as well as their coexistence with and evolution to other quantum orders. Initially, the correlated metallic transition-metal dichalcogenide (TMD) received attention for its four distinct charge density wave (CDW) phases.\cite{Nakanishi1977, Tanda1984, Thomson1994} 1T-TaS$_2$ has also garnered extensive interest for a potential low-temperature Mott insulating state coexisting with the low-temperature CDW state.\cite{Law2017,Cho2016,Ma2016,Perfetti2006,Butler2020} 

The proximity of multiple quantum orders in 1T-TaS$_2$ has led to an interest in tuning these electron-electron and electron-phonon interactions to alter and control the material’s electronic and structural properties. Elemental doping has led to the emergence of superconductivity \cite{Liu2013,Li2012} and to the reported evolution of the Mott insulating state to a metallic state \cite{Qiao2017}. Perturbations induced by electromagnetic fields and strain have led to new stable electronic states.\cite{Ma2016,Stojchevska2014,Vaskivskyi2015,Gerasimenko2019,Bu2019} The application of pressure has led to the emergence of superconductivity and coexistence of the superconducting state with a CDW state.\cite{Ritschel2013,Sipos2008} Altering the thickness of 1T-TaS$_2$ crystals provides another mechanism by which to control the 1T-TaS$_2$ phase diagram.\cite{Yoshida2014,Yoshida2015} Emerging from these studies is a better understanding not only of the fundamental physics hosted by the 1T-TaS$_2$ but also of the mechanisms by which to appropriately tailor the material’s properties for possible future applications in electronics and data storage.

\begin{figure}[ht]
\includegraphics[clip=true,width=\columnwidth]{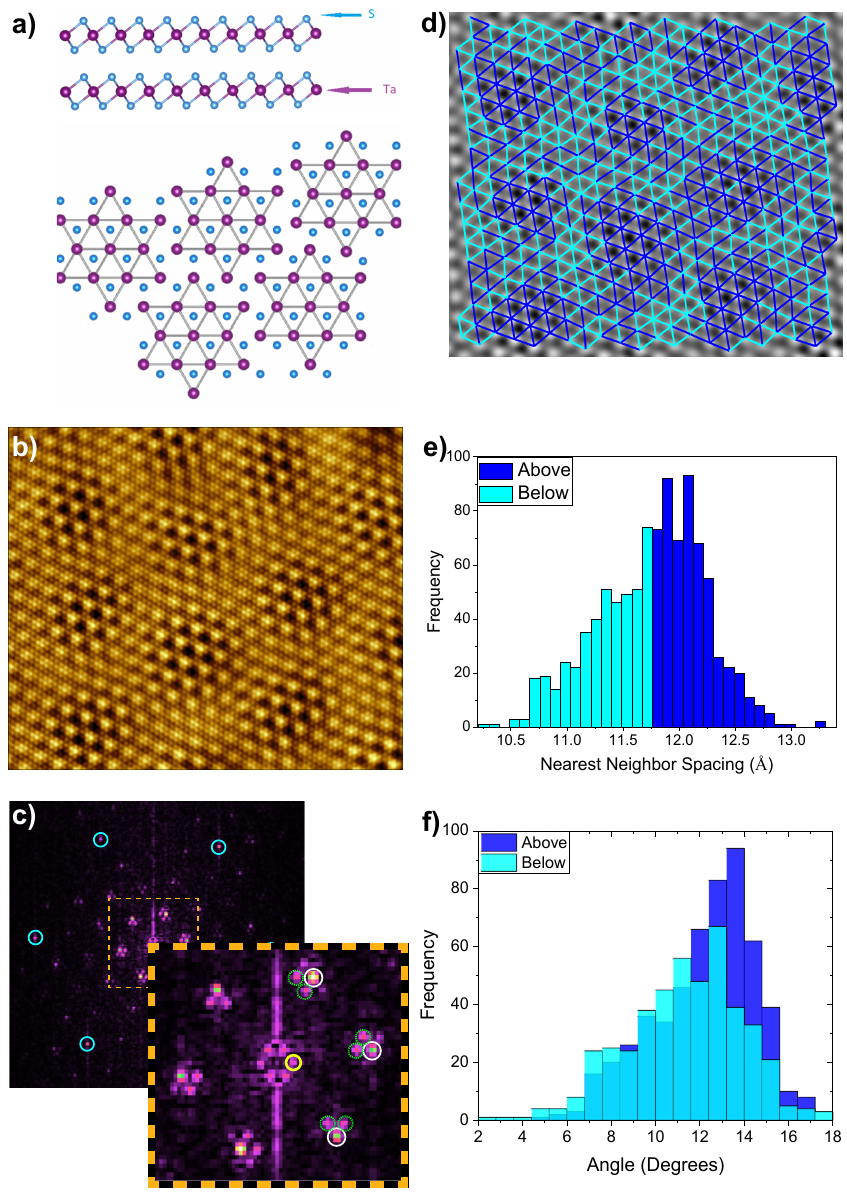}
\caption{a) Top: Layered crystal structure for 1T-TaS$_2$. Bottom: Top view showing Ta-formed David stars. Surface S ions, directly imaged by STM, are superimposed for reference. Crystal structure was created using Vesta.\cite{Momma2011} b) 230 {\AA} x 200 {\AA} STM topography taken parameters I =  450 pA and V$_Bias$ = 150 mV. Image was Fourier filtered to include signals identified in (c). c) Fourier transform of 230 {\AA} x 200 {\AA} topographic image at left. Atomic signals are circled in blue. Large orange dotted square is zoom of central FFT region. Within this central region are low wavevector peaks (yellow) and average CDW peaks (white) around which there are satellite peaks (dotted green). d) Connections between NN CDW maxima superimposed on Fourier filtered topographic image (now grayscale). Filtered image shows only locations of CDW maxima. Average NN distance is 11.76 {\AA}. Blue (cyan) connections show NN separations above (below) 11.76 {\AA} threshold. e) Histogram of NN distances. f) Angles of NN connectors relative to atomic lattice. Blue (cyan) show angular distributions for connectors above (below) 11.76 {\AA} threshold.}
\label{fig:Fig1}
\end{figure}

Here we focus on our scanning tunneling microscopy (STM) studies of the nearly-commensurate CDW (NC-CDW) state of 1T-TaS$_2$. Figure \ref{fig:Fig1}a shows the crystal structure for the undistorted material.  At a temperature of $\sim$543 K, the Ta ions in the S-Ta-S layers distort and 1T-TaS$_2$ transitions from a metallic to an incommensurate CDW (I-CDW) state.\cite{Thomson1994} At temperatures below $\sim$180 K, 1T-TaS$_2$ enters a commensurate CDW (C-CDW) state. In this CDW state, 12 Ta ions distort toward a central Ta ion in a ‘David star’-like pattern.\cite{Ishiguro1991} These stars then create a $\sqrt{13}$a$_0$×$\sqrt{13}$a$_0$ commensurate pattern across the layer. 

Between the I-CDW and C-CDW phases resides the NC-CDW state. The transition from the IC-CDW to the NC-CDW state occurs at $\sim$350 K.\cite{Scruby1975} The NC-CDW state is commonly characterized as having hexagonally ordered C-CDW domains separated by an IC-CDW domain-wall network or by discommensurations.\cite{Ma2016,Yoshida2015,Shao2016,Wang2019} However, as has previously been noted \cite{Thomson1994}, this description may be a simplification of the true NC-CDW domain structure. Lacking is a characterization of the NC-CDW state which details with sub-angstrom resolution the evolution of the CDW state from the domain wall to the domain center. Given that distortions of Ta ions within 1T-TaS$_2$ are fundamental to driving the observed CDW states and electronic properties of the material, there is an essential need to better-characterize and understand the nanoscale modulations of the NC-CDW state from the center of domains to the domain walls. Further, as we characterize the NC state in 1T-TaS$_2$, we uncover a previously unreported intradomain CDW chirality.

Figure \ref{fig:Fig1}b shows an STM topographic image of the NC-CDW state we acquired at room temperature. The acquired-image shows a roughly hexagonally-ordered domain pattern characteristic of the NC state. The fast Fourier transform (FFT) of the image (Figure \ref{fig:Fig1}c) evinces essential peaks from which we extract average quantities characterizing the NC-CDW state. We find the CDW superlattice to have an average periodicity of 11.76 {\AA} at an average angle of 11.8{\degree} relative to the atomic lattice, in agreement with previous STM, x-ray, and electron diffraction measurements.\cite{Nakanishi1977,Thomson1994,Ishiguro1991} Using the position of satellite peaks relative to the CDW peaks allows us to determine the average domain periodicity of 73 {\AA}, also in agreement.\cite{Thomson1994,Tsen2015}

To fully understand the NC-CDW state requires going beyond average quantities. To do this, we focus only on the CDW superlattice and Fourier filter our topographic image such that only the central CDW and surrounding satellite peaks are included (Figure \ref{fig:Fig1}d). We next calculate nearest neighbor (NN) distances between CDW maxima so as to understand the distribution of distances. We find a large range of NN distances spanning 10.26 {\AA} to 13.22 {\AA}. Superimposed on the image in Figure \ref{fig:Fig1}d are line segments connecting the CDW maxima which are color coded according to the 11.76 {\AA} threshold with segments colored dark blue (cyan) being greater (smaller) than the threshold. Generally, the CDW maxima have a greater spacing within the domain than in the domain walls.

In the NC-CDW state, the CDW lattice is rotated relative to the atomic lattice. This is seen in Figure \ref{fig:Fig1}c where the CDW peaks are rotated by 11.8{\degree} relative to the atomic peaks in the FFT, in agreement with angles as measured by electron-diffraction [20] and STM measurements [3] acquired at this temperature. Using the NN distance threshold of 11.76 {\AA}, we can create histograms for the local CDW lattice rotation relative to the atomic lattice for NNs above and below the threshold as seen in Figures \ref{fig:Fig1}e and \ref{fig:Fig1}f. While the average relative angle for NN above the distance threshold (those generally within a domain) of 12.1{\degree} is greater than the average for those below (those generally within the domain wall) of 11.3{\degree}, there is a wide range of angles within each grouping.

\begin{figure}[t]
\includegraphics[clip=true,width=\columnwidth]{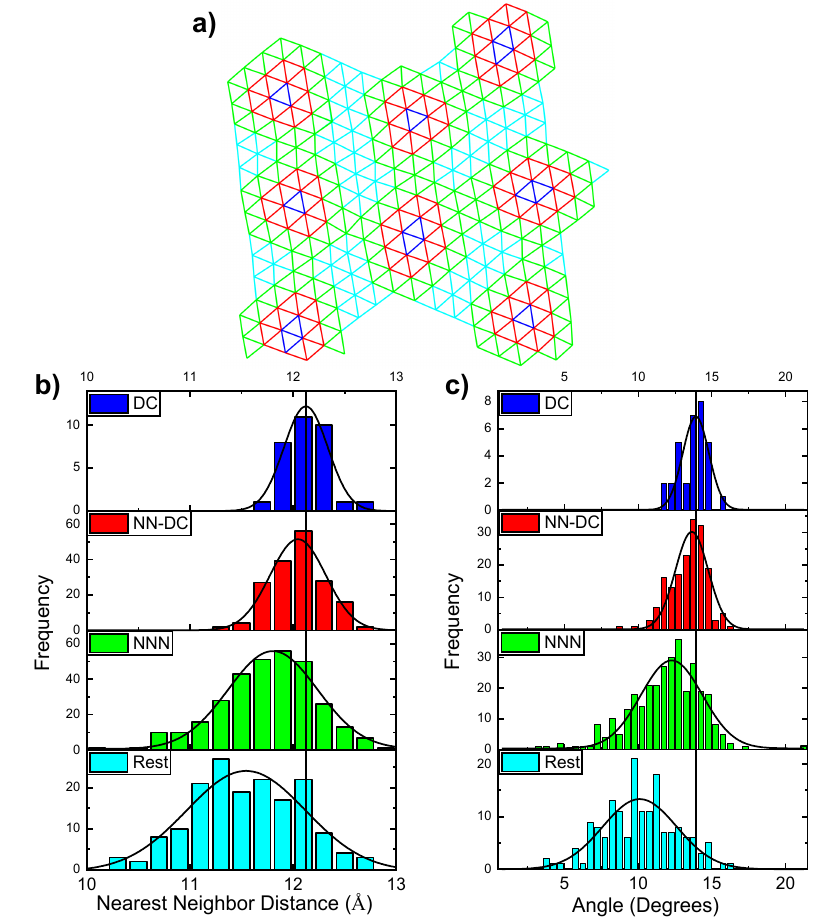}
\caption{a) Domain center (blue) found using local intensity thresholding – here 95{\%} of most intense CDW maximum within a domain define center. NN-DC (red) nearest neighbor CDW connectors towards domain center (excludes blue connectors). NNN – next nearest neighbor connectors towards domain center (excludes blue and red). Rest – remaining connectors (excludes blue, red, cyan). Domains with centers near the edge of the topography are ignored. b) Histograms of NN distances for each connector grouping. c) Histograms of CDW angles relative to atomic lattice for each connector grouping. Gaussian fits and vertical line at DC peak are superimposed in b) and c) to illustrate evolution of peak location and width.}
\label{fig:Fig2}
\end{figure}

To obtain a better idea of the evolution of the CDW spacing and angles from domain wall to domain center, we employ a local masking technique. For a given domain, we identify all CDW maxima with an intensity which is within 95{\%} of the maximum CDW intensity of that domain. We then calculate the NN distances among each of these CDW peaks thereby quantifying NN distances at the center of the domain (Figure \ref{fig:Fig2}a blue). We next include the NN CDW peaks to the domain center CDW peaks and recalculate NN distances omitting the NN domain center values (Figure \ref{fig:Fig2}a red). We then include the next nearest neighbor (NNN) peaks (Figure \ref{fig:Fig2}a green). Finally, we include all remaining CDW peaks (Figure \ref{fig:Fig2}a cyan) allowing us to determine the evolution of the NN distances progressing from domain walls to domain centers (Figure \ref{fig:Fig2}b). Obvious from the histograms are: 1) an increase in NN spacing and 2) a narrowing of the distribution width as one moves from the domain walls toward the domain center. These two apparent progressions are confirmed with the evolution of the mean and standard deviation of 11.54${\pm}$0.52 {\AA} (cyan) to 11.75${\pm}$0.49 {\AA} (green) to 12.06${\pm}$0.27 {\AA} (red) to 12.14${\pm}$0.20 {\AA} (blue). We note that the NN spacing in our measurements directly at the domain center is slightly larger than the fixed NN CDW spacing of the low-temperature C-CDW state. Using the same groupings, we determine the local CDW lattice angle relative to the atomic lattice. There is a similar evolution in the mean and distribution width of the CDW angle toward the low-temperature commensurate value of 13.9{\degree} as one progresses from the domain walls to the domain centers (Figure \ref{fig:Fig2}c): 10.1{\degree}${\pm}$2.5{\degree} (cyan) to 11.9{\degree}${\pm}$2.4{\degree} (green) to 13.4{\degree}${\pm}$1.2{\degree} (red) to 13.7{\degree}${\pm}$0.9{\degree} (blue).

\begin{figure}[b]
\includegraphics[clip=true,width=\columnwidth]{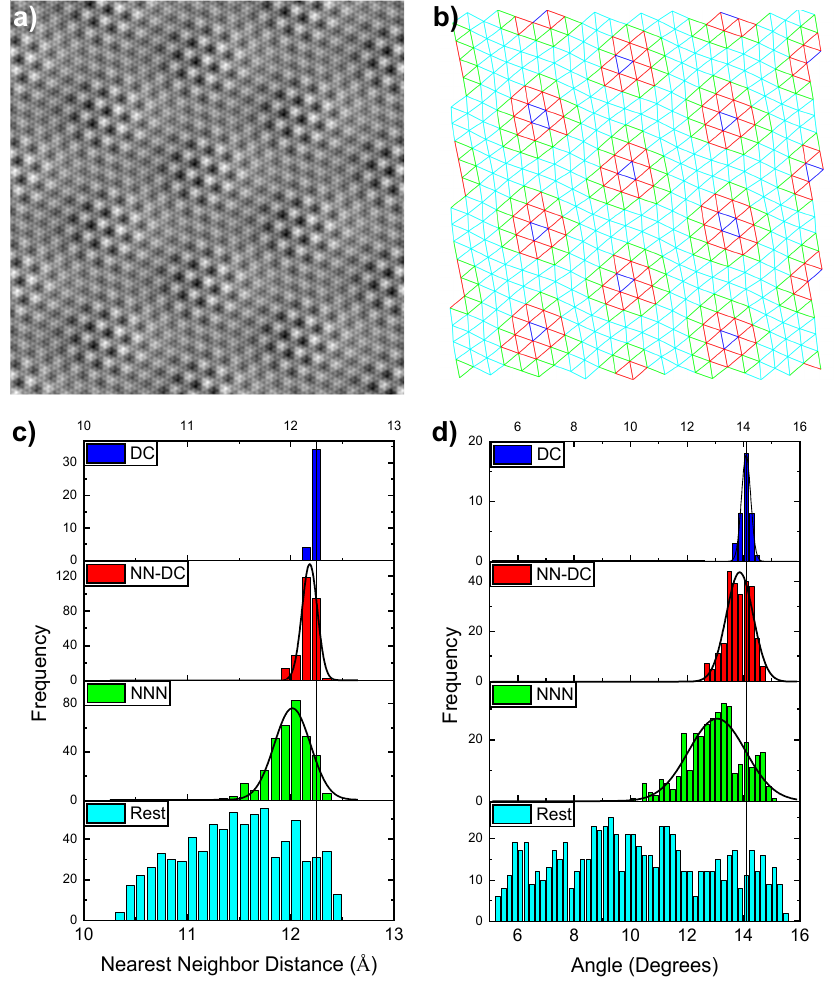}
\caption{a) 253 {\AA} x 253 {\AA} simulated STM topography. Average intensity and wavevectors for atomic, CDW, satellite, and low-wavevector peaks from Figure \ref{fig:Fig1}c were used to create image. Note, Figure \ref{fig:Fig1}c and other data we acquire show two satellite peaks which are more intense than others. The simulated STM topography includes only signals from the two most intense satellite peaks around a given CDW peak.  b) Domain center (blue) found using local intensity thresholding. NN-DC (red) nearest neighbor CDW connectors towards domain center (excludes blue connectors). NNN – next nearest neighbor connectors towards domain center (excludes blue and red). Rest – remaining connectors (excludes blue, red, cyan). c) Histograms of NN distances for each connector grouping. Statistics: 11.48${\pm}$0.53 {\AA} (cyan) to 11.98${\pm}$0.19 {\AA} (green) to 12.16${\pm}$0.08 {\AA} (red) to 12.23${\pm}$0.03 {\AA} (blue) d) Histograms of CDW angles relative to atomic lattice for each connector grouping. Statistics: 10.2{\degree}${\pm}$2.7{\degree} (cyan) to 13.0{\degree}${\pm}$1.0{\degree} (green) to 13.8{\degree}${\pm}$0.5{\degree} (red) to 14.1{\degree}${\pm}$0.2{\degree} (blue). Gaussian fits and vertical line at DC peak are superimposed in c) and d) to illustrate evolution of peak location and width.}
\label{fig:Fig3}
\end{figure}

Underlying crystal disorder and defects within a sample can obscure fundamental/overarching characteristics within a material. To obtain a clearer progression of the NN distances from the domain walls to the domain center, we used the wavevectors extracted from the FFT (Figure \ref{fig:Fig1}c) to simulate our topographic image (Figure \ref{fig:Fig3}a). Our simulated topography well-matches our topographic image in Figure \ref{fig:Fig1}b. Using intensity thresholding we calculate NN distances and angles starting at the domain centers and progressing toward the domain walls (Figure \ref{fig:Fig3}b). The histograms appearing in Figures \ref{fig:Fig3}c and \ref{fig:Fig3}d illustrate the range of values and confirm the progression found in the data: the average NN CDW spacing and angle continually increase from domain wall to domain center, and the width of the distributions narrows only approaching values found for the low-temperature C-CDW state near the center of a domain.

\begin{figure}[ht]
\includegraphics[clip=true,width=\columnwidth]{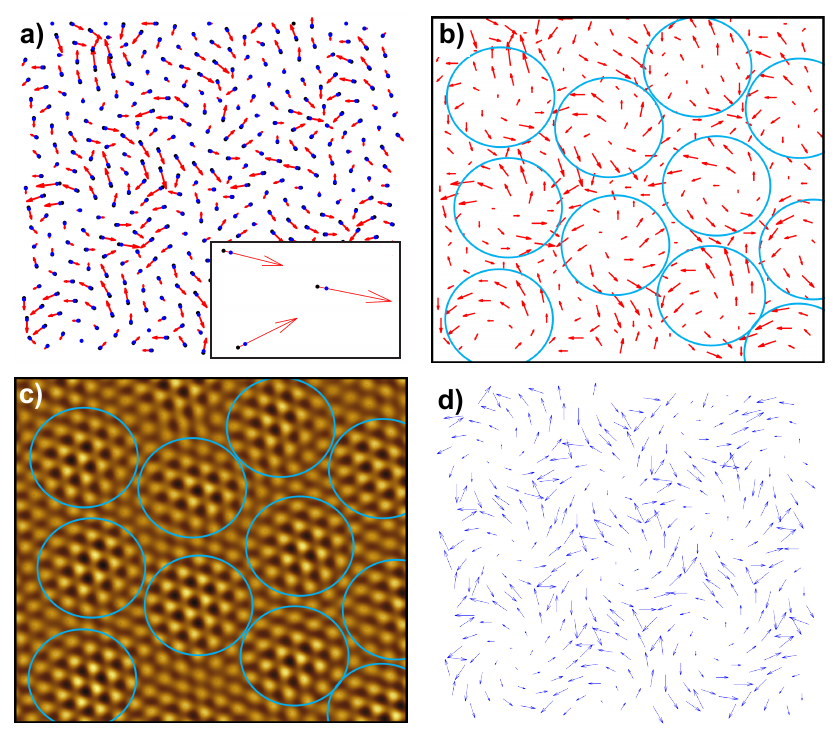}
\caption{a) Arrows point from average CDW maximum location to true CDW maximum location. Arrow length is proportional to displacement. Inset shows zoom and illustrates arrows pointing from average CDW maximum (black) to true CDW maximum (blue). b) Blue circles are superimposed on arrows from a) as a guide to the eye to show the vortex-like chiral displacement patterns. c) The same blue circles are superimposed on topography (with atomic signal removed) and match domain locations. d) Arrows are drawn from the average CDW maximum location to the nearest imaged sulfur ion. An identical, though translated, pattern is produced if arrows are drawn from the average CDW maximum location to the nearest Ta ions. We chose the sulfur ion for this figure since we image those ions directly and obtain the same information.}
\label{fig:Fig4}
\end{figure}

Having elucidated the spacing and local angle of the CDW evolution from domain wall to domain center, we study the connection satellite peaks (Figure \ref{fig:Fig1}c) have to the domain structure observed in topographic images. Fourier filtering of the topographic image (Figure \ref{fig:Fig1}b) where only the central CDW peaks in the FFT are included (white circles in Figure \ref{fig:Fig1}c) produces a topography with purely hexagonal CDW lattice with NN CDW spacing (and angle relative to the lattice) equal to the average CDW periodicity (and angle) for the material, and which is purely incommensurate with the atomic lattice. We will call this the “average CDW” lattice. Including the satellite peaks (in addition to the central CDW peak when Fourier filtering) leads to a CDW with domain structure as seen in Figure \ref{fig:Fig1}d. We will call this the “true CDW” lattice. We now compare the two images. Figure \ref{fig:Fig4}a shows the center locations for CDW maxima for 1) average CDW (black dots) and 2) true CDW (blue dots). Arrows superimposed on this image point from the average to the true CDW location with the length of the arrow proportional to the shift distance.

The shift of the true CDW peak location relative to its average CDW peak location varies spatially in a way that mimics the domain pattern of the NC-CDW state. Further, the shift arrows can be used to easily identify the locations and extent of individual domains. Within a domain, each shift arrow has a continual evolution in direction (counterclockwise rotation) and size (the shift is smallest at the domain center and largest at the domain edge). In short, we uncover a previously unreported intradomain chirality to the NC-CDW state. At the domain wall, there is a sudden shift in this evolution with shift arrows from neighboring domains sometimes pointing in direct opposition to one another. This shift is indicative of the CDW discommensuration between domains. Using the shift arrows as reference, we overlay blue circles as a general guide to the eye to illustrate the domain locations, rough size, and arrangement (Figure \ref{fig:Fig4}b). The same circles are superimposed on the true CDW filtered image (Figure \ref{fig:Fig4}c) verifying the ability to use the shift arrows to appropriately identify domain locations. Quantifying the true CDW shift from the average CDW maximum location, we find an average displacement of 0.72 {\AA} with a maximum displacement of 1.70 {\AA}.  This maximum displacement is significant and represents a displacement of $\sim$14$\%$ of the average CDW lattice spacing. Simulating the average and true CDW lattices confirms and provides even clearer evidence of this chirality (supplemental materials).

The observed shifts highlight an intrinsic chirality to the NC-CDW state: there is a counter-clockwise vortex-like displacement of the true CDW maxima relative to the average CDW locations. The signature of this intradomain chirality is embedded in the satellite peaks in the FFT of the topographic image. While Figure \ref{fig:Fig1}c show multiple satellite peaks around several of the CDW peaks in the FFT, each CDW peak (white circles) has two associated high-intensity satellite peaks (green dotted circles). These two intense satellite peaks rotate clockwise around the CDW peaks as one progresses clockwise around the FFT center. FFTs of our topographic images of the NC-CDW state indeed commonly show two intense satellite peaks which rotate around an average CDW peak and is also commonly seen in other published STM studies of the NC-CDW state of 1T-TaS$_2$.\cite{Thomson1994,Gerasimenko2019}

Finally, we explore the origin of the chiral nature of the CDW state. A chiral CDW state has been observed in 1T-TiSe$_2$ \cite{Ishioka2010} and recently in the low-temperature state of Ti-doped 1T-TaS$_2$ \cite{Gao2021}. The chiral nature these states originates from the orbital structure.\cite{Gao2021,Klosinski2021,vanWezel2011,Peng2021} The chirality is detected in FFTs of STM topographies through CDW peak intensities which increase/decrease clockwise or anticlockwise.\cite{Ishioka2010,Gao2021} We detect no such intensity progression in the CDW peak intensities in our studies of the NC-CDW state of 1T-TaS$_2$. Rather, in our FFTs we find a clockwise rotation of two intense satellite peaks around the central CDW peaks, connected to the domain formation and chiral CDW lattice distortions from the average CDW lattice. Figure \ref{fig:Fig4}d shows arrows pointing from the average CDW maxima to the nearest imaged atom evincing a very similar chiral pattern to Figure \ref{fig:Fig4}a. In each, shifts are in the counterclockwise direction leading to clear domain formation. In short, Figure \ref{fig:Fig4}d shows the CDW displacements in the limit of a perfectly incommensurate CDW lattice which has very strong homogenous coupling to the atomic lattice with no energy cost for CDW deformations. This strongly suggests that the chiral domains formed in the NC-CDW state of 1T-TaS$_2$ are driven through a strong coupling of the CDW state to the atomic lattice, not through orbital ordering which is believed to drive chiral CDW orders in related transition-metal dichalcogenides. We can understand the origin of the strong coupling of the CDW to the lattice in the NC-CDW state of 1T-TaS$_2$ in the following way. Within a David star, there are 13 5d-electrons, one from each Ta ion, 12 of which are bonded leaving a single “extra” electron per star.\cite{Kratochvilova2017} The strong coupling of the CDW state to the lattice likely originates from the Coulomb interaction of these extra electrons with the Ta lattice; these extra electrons preferentially locate near Ta ions.

Previous chiral CDW orders found in the transition metal dichalcogenides (TMDCs) have been attributed to orbital ordering. In the NC-CDW state of the TMDC 1T-TaS$_2$ we find, instead, a CDW state with chiral domains which appears to originate from a strong coupling of an incommensurate CDW to the lattice. Such domain chirality may be a more-general phenomena which can be explored in systems hosting incommensurate CDW states, including in TMDCs.

\subsection{Methods}
STM measurements were conducted at room temperature and ulra-high vacuum ($\sim$10$^{-9}$ Torr) using a tungsten tip which was chemically etched. The tip was annealed and then sharpened using electron bombardment in-situ. High-quality single crystals of 1T-TaS$_2$ studied in this were purchased from HQ Graphene. Samples were cleaved in UHV.

\subsection{Acknowledgements}
This work was supported by the National Science Foundation under Grant No. DMR-1904918.


\bibliography{TaS2.bib}  

\begin{thebibliography}{31}%
\makeatletter
\providecommand \@ifxundefined [1]{%
 \@ifx{#1\undefined}
}%
\providecommand \@ifnum [1]{%
 \ifnum #1\expandafter \@firstoftwo
 \else \expandafter \@secondoftwo
 \fi
}%
\providecommand \@ifx [1]{%
 \ifx #1\expandafter \@firstoftwo
 \else \expandafter \@secondoftwo
 \fi
}%
\providecommand \natexlab [1]{#1}%
\providecommand \enquote  [1]{``#1''}%
\providecommand \bibnamefont  [1]{#1}%
\providecommand \bibfnamefont [1]{#1}%
\providecommand \citenamefont [1]{#1}%
\providecommand \href@noop [0]{\@secondoftwo}%
\providecommand \href [0]{\begingroup \@sanitize@url \@href}%
\providecommand \@href[1]{\@@startlink{#1}\@@href}%
\providecommand \@@href[1]{\endgroup#1\@@endlink}%
\providecommand \@sanitize@url [0]{\catcode `\\12\catcode `\$12\catcode
  `\&12\catcode `\#12\catcode `\^12\catcode `\_12\catcode `\%12\relax}%
\providecommand \@@startlink[1]{}%
\providecommand \@@endlink[0]{}%
\providecommand \url  [0]{\begingroup\@sanitize@url \@url }%
\providecommand \@url [1]{\endgroup\@href {#1}{\urlprefix }}%
\providecommand \urlprefix  [0]{URL }%
\providecommand \Eprint [0]{\href }%
\providecommand \doibase [0]{https://doi.org/}%
\providecommand \selectlanguage [0]{\@gobble}%
\providecommand \bibinfo  [0]{\@secondoftwo}%
\providecommand \bibfield  [0]{\@secondoftwo}%
\providecommand \translation [1]{[#1]}%
\providecommand \BibitemOpen [0]{}%
\providecommand \bibitemStop [0]{}%
\providecommand \bibitemNoStop [0]{.\EOS\space}%
\providecommand \EOS [0]{\spacefactor3000\relax}%
\providecommand \BibitemShut  [1]{\csname bibitem#1\endcsname}%
\let\auto@bib@innerbib\@empty
\bibitem [{\citenamefont {Nakanishi}\ \emph {et~al.}(1977)\citenamefont
  {Nakanishi}, \citenamefont {Takatera}, \citenamefont {Yamada},\ and\
  \citenamefont {Shiba}}]{Nakanishi1977}%
  \BibitemOpen
  \bibfield  {author} {\bibinfo {author} {\bibfnamefont {K.}~\bibnamefont
  {Nakanishi}}, \bibinfo {author} {\bibfnamefont {H.}~\bibnamefont {Takatera}},
  \bibinfo {author} {\bibfnamefont {Y.}~\bibnamefont {Yamada}},\ and\ \bibinfo
  {author} {\bibfnamefont {H.}~\bibnamefont {Shiba}},\ }\bibfield  {title}
  {\bibinfo {title} {The nearly commensurate phase and effect of harmonics on
  the successive phase transition in 1\ch{T-TaS$_2$}},\ }\href
  {https://doi.org/10.1143/JPSJ.43.1509} {\bibfield  {journal} {\bibinfo
  {journal} {Journal of the Physical Society of Japan}\ }\textbf {\bibinfo
  {volume} {43}},\ \bibinfo {pages} {1509} (\bibinfo {year}
  {1977})}\BibitemShut {NoStop}%
\bibitem [{\citenamefont {Tanda}\ \emph {et~al.}(1984)\citenamefont {Tanda},
  \citenamefont {Sambongi}, \citenamefont {Tani},\ and\ \citenamefont
  {Tanaka}}]{Tanda1984}%
  \BibitemOpen
  \bibfield  {author} {\bibinfo {author} {\bibfnamefont {S.}~\bibnamefont
  {Tanda}}, \bibinfo {author} {\bibfnamefont {T.}~\bibnamefont {Sambongi}},
  \bibinfo {author} {\bibfnamefont {T.}~\bibnamefont {Tani}},\ and\ \bibinfo
  {author} {\bibfnamefont {S.}~\bibnamefont {Tanaka}},\ }\bibfield  {title}
  {\bibinfo {title} {X-ray study of charge density wave structure in
  1\ch{T-TaS$_2$}},\ }\href {https://doi.org/10.1143/JPSJ.53.476} {\bibfield
  {journal} {\bibinfo  {journal} {Journal of the Physical Society of Japan}\
  }\textbf {\bibinfo {volume} {53}},\ \bibinfo {pages} {476} (\bibinfo {year}
  {1984})}\BibitemShut {NoStop}%
\bibitem [{\citenamefont {Thomson}\ \emph {et~al.}(1994)\citenamefont
  {Thomson}, \citenamefont {Burk}, \citenamefont {Zettl},\ and\ \citenamefont
  {Clarke}}]{Thomson1994}%
  \BibitemOpen
  \bibfield  {author} {\bibinfo {author} {\bibfnamefont {R.~E.}\ \bibnamefont
  {Thomson}}, \bibinfo {author} {\bibfnamefont {B.}~\bibnamefont {Burk}},
  \bibinfo {author} {\bibfnamefont {A.}~\bibnamefont {Zettl}},\ and\ \bibinfo
  {author} {\bibfnamefont {J.}~\bibnamefont {Clarke}},\ }\bibfield  {title}
  {\bibinfo {title} {Scanning tunneling microscopy of the charge-density-wave
  structure in 1\ch{T-TaS$_2$}},\ }\href
  {https://doi.org/10.1103/PhysRevB.49.16899} {\bibfield  {journal} {\bibinfo
  {journal} {Physical Review B}\ }\textbf {\bibinfo {volume} {49}},\ \bibinfo
  {pages} {16899} (\bibinfo {year} {1994})}\BibitemShut {NoStop}%
\bibitem [{\citenamefont {Law}\ and\ \citenamefont {Lee}(2017)}]{Law2017}%
  \BibitemOpen
  \bibfield  {author} {\bibinfo {author} {\bibfnamefont {K.~T.}\ \bibnamefont
  {Law}}\ and\ \bibinfo {author} {\bibfnamefont {P.~A.}\ \bibnamefont {Lee}},\
  }\bibfield  {title} {\bibinfo {title} {1\ch{T-TaS2} as a quantum spin
  liquid},\ }\href {https://doi.org/10.1073/pnas.1706769114} {\bibfield
  {journal} {\bibinfo  {journal} {Proceedings of the National Academy of
  Sciences}\ }\textbf {\bibinfo {volume} {114}},\ \bibinfo {pages} {6996}
  (\bibinfo {year} {2017})}\BibitemShut {NoStop}%
\bibitem [{\citenamefont {Cho}\ \emph {et~al.}(2016)\citenamefont {Cho},
  \citenamefont {Cheon}, \citenamefont {Kim}, \citenamefont {Lee},
  \citenamefont {Cho}, \citenamefont {Cheong},\ and\ \citenamefont
  {Yeom}}]{Cho2016}%
  \BibitemOpen
  \bibfield  {author} {\bibinfo {author} {\bibfnamefont {D.}~\bibnamefont
  {Cho}}, \bibinfo {author} {\bibfnamefont {S.}~\bibnamefont {Cheon}}, \bibinfo
  {author} {\bibfnamefont {K.-S.}\ \bibnamefont {Kim}}, \bibinfo {author}
  {\bibfnamefont {S.-H.}\ \bibnamefont {Lee}}, \bibinfo {author} {\bibfnamefont
  {Y.-H.}\ \bibnamefont {Cho}}, \bibinfo {author} {\bibfnamefont {S.-W.}\
  \bibnamefont {Cheong}},\ and\ \bibinfo {author} {\bibfnamefont {H.~W.}\
  \bibnamefont {Yeom}},\ }\bibfield  {title} {\bibinfo {title} {Nanoscale
  manipulation of the \ch{M}ott insulating state coupled to charge order in
  1\ch{T}-\ch{TaS2}},\ }\href {https://doi.org/10.1038/ncomms10453} {\bibfield
  {journal} {\bibinfo  {journal} {Nature Communications}\ }\textbf {\bibinfo
  {volume} {7}},\ \bibinfo {pages} {10453} (\bibinfo {year}
  {2016})}\BibitemShut {NoStop}%
\bibitem [{\citenamefont {Ma}\ \emph {et~al.}(2016)\citenamefont {Ma},
  \citenamefont {Ye}, \citenamefont {Yu}, \citenamefont {Lu}, \citenamefont
  {Niu}, \citenamefont {Kim}, \citenamefont {Feng}, \citenamefont {Tománek},
  \citenamefont {Son}, \citenamefont {Chen},\ and\ \citenamefont
  {Zhang}}]{Ma2016}%
  \BibitemOpen
  \bibfield  {author} {\bibinfo {author} {\bibfnamefont {L.}~\bibnamefont
  {Ma}}, \bibinfo {author} {\bibfnamefont {C.}~\bibnamefont {Ye}}, \bibinfo
  {author} {\bibfnamefont {Y.}~\bibnamefont {Yu}}, \bibinfo {author}
  {\bibfnamefont {X.~F.}\ \bibnamefont {Lu}}, \bibinfo {author} {\bibfnamefont
  {X.}~\bibnamefont {Niu}}, \bibinfo {author} {\bibfnamefont {S.}~\bibnamefont
  {Kim}}, \bibinfo {author} {\bibfnamefont {D.}~\bibnamefont {Feng}}, \bibinfo
  {author} {\bibfnamefont {D.}~\bibnamefont {Tománek}}, \bibinfo {author}
  {\bibfnamefont {Y.-W.}\ \bibnamefont {Son}}, \bibinfo {author} {\bibfnamefont
  {X.~H.}\ \bibnamefont {Chen}},\ and\ \bibinfo {author} {\bibfnamefont
  {Y.}~\bibnamefont {Zhang}},\ }\bibfield  {title} {\bibinfo {title} {A
  metallic mosaic phase and the origin of \ch{M}ott-insulating state in
  1\ch{T-TaS$_2$}},\ }\href {https://doi.org/10.1038/ncomms10956} {\bibfield
  {journal} {\bibinfo  {journal} {Nature Communications}\ }\textbf {\bibinfo
  {volume} {7}},\ \bibinfo {pages} {10956} (\bibinfo {year}
  {2016})}\BibitemShut {NoStop}%
\bibitem [{\citenamefont {Perfetti}\ \emph {et~al.}(2006)\citenamefont
  {Perfetti}, \citenamefont {Loukakos}, \citenamefont {Lisowski}, \citenamefont
  {Bovensiepen}, \citenamefont {Berger}, \citenamefont {Biermann},
  \citenamefont {Cornaglia}, \citenamefont {Georges},\ and\ \citenamefont
  {Wolf}}]{Perfetti2006}%
  \BibitemOpen
  \bibfield  {author} {\bibinfo {author} {\bibfnamefont {L.}~\bibnamefont
  {Perfetti}}, \bibinfo {author} {\bibfnamefont {P.~A.}\ \bibnamefont
  {Loukakos}}, \bibinfo {author} {\bibfnamefont {M.}~\bibnamefont {Lisowski}},
  \bibinfo {author} {\bibfnamefont {U.}~\bibnamefont {Bovensiepen}}, \bibinfo
  {author} {\bibfnamefont {H.}~\bibnamefont {Berger}}, \bibinfo {author}
  {\bibfnamefont {S.}~\bibnamefont {Biermann}}, \bibinfo {author}
  {\bibfnamefont {P.~S.}\ \bibnamefont {Cornaglia}}, \bibinfo {author}
  {\bibfnamefont {A.}~\bibnamefont {Georges}},\ and\ \bibinfo {author}
  {\bibfnamefont {M.}~\bibnamefont {Wolf}},\ }\bibfield  {title} {\bibinfo
  {title} {Time evolution of the electronic structure of 1\ch{T-TaS$_2$}
  through the insulator-metal transition},\ }\href
  {https://doi.org/10.1103/PhysRevLett.97.067402} {\bibfield  {journal}
  {\bibinfo  {journal} {Physical Review Letters}\ }\textbf {\bibinfo {volume}
  {97}},\ \bibinfo {pages} {067402} (\bibinfo {year} {2006})}\BibitemShut
  {NoStop}%
\bibitem [{\citenamefont {Butler}\ \emph {et~al.}(2020)\citenamefont {Butler},
  \citenamefont {Yoshida}, \citenamefont {Hanaguri},\ and\ \citenamefont
  {Iwasa}}]{Butler2020}%
  \BibitemOpen
  \bibfield  {author} {\bibinfo {author} {\bibfnamefont {C.~J.}\ \bibnamefont
  {Butler}}, \bibinfo {author} {\bibfnamefont {M.}~\bibnamefont {Yoshida}},
  \bibinfo {author} {\bibfnamefont {T.}~\bibnamefont {Hanaguri}},\ and\
  \bibinfo {author} {\bibfnamefont {Y.}~\bibnamefont {Iwasa}},\ }\bibfield
  {title} {\bibinfo {title} {Mottness versus unit-cell doubling as the driver
  of the insulating state in 1\ch{T-TaS2}},\ }\href
  {https://doi.org/10.1038/s41467-020-16132-9} {\bibfield  {journal} {\bibinfo
  {journal} {Nature Communications}\ }\textbf {\bibinfo {volume} {11}},\
  \bibinfo {pages} {2477} (\bibinfo {year} {2020})}\BibitemShut {NoStop}%
\bibitem [{\citenamefont {Liu}\ \emph {et~al.}(2013)\citenamefont {Liu},
  \citenamefont {Ang}, \citenamefont {Lu}, \citenamefont {Song}, \citenamefont
  {Li},\ and\ \citenamefont {Sun}}]{Liu2013}%
  \BibitemOpen
  \bibfield  {author} {\bibinfo {author} {\bibfnamefont {Y.}~\bibnamefont
  {Liu}}, \bibinfo {author} {\bibfnamefont {R.}~\bibnamefont {Ang}}, \bibinfo
  {author} {\bibfnamefont {W.~J.}\ \bibnamefont {Lu}}, \bibinfo {author}
  {\bibfnamefont {W.~H.}\ \bibnamefont {Song}}, \bibinfo {author}
  {\bibfnamefont {L.~J.}\ \bibnamefont {Li}},\ and\ \bibinfo {author}
  {\bibfnamefont {Y.~P.}\ \bibnamefont {Sun}},\ }\bibfield  {title} {\bibinfo
  {title} {Superconductivity induced by \ch{Se}-doping in layered
  charge-density-wave system 1\ch{T-TaS$_{2-x}$Se$_x$}},\ }\href
  {https://doi.org/10.1063/1.4805003} {\bibfield  {journal} {\bibinfo
  {journal} {Applied Physics Letters}\ }\textbf {\bibinfo {volume} {102}},\
  \bibinfo {pages} {192602} (\bibinfo {year} {2013})}\BibitemShut {NoStop}%
\bibitem [{\citenamefont {Li}\ \emph {et~al.}(2012)\citenamefont {Li},
  \citenamefont {Lu}, \citenamefont {Zhu}, \citenamefont {Ling}, \citenamefont
  {Qu},\ and\ \citenamefont {Sun}}]{Li2012}%
  \BibitemOpen
  \bibfield  {author} {\bibinfo {author} {\bibfnamefont {L.~J.}\ \bibnamefont
  {Li}}, \bibinfo {author} {\bibfnamefont {W.~J.}\ \bibnamefont {Lu}}, \bibinfo
  {author} {\bibfnamefont {X.~D.}\ \bibnamefont {Zhu}}, \bibinfo {author}
  {\bibfnamefont {L.~S.}\ \bibnamefont {Ling}}, \bibinfo {author}
  {\bibfnamefont {Z.}~\bibnamefont {Qu}},\ and\ \bibinfo {author}
  {\bibfnamefont {Y.~P.}\ \bibnamefont {Sun}},\ }\bibfield  {title} {\bibinfo
  {title} {Fe-doping–induced superconductivity in the charge-density-wave
  system 1\ch{T-TaS2}},\ }\href {https://doi.org/10.1209/0295-5075/97/67005}
  {\bibfield  {journal} {\bibinfo  {journal} {EPL (Europhysics Letters)}\
  }\textbf {\bibinfo {volume} {97}},\ \bibinfo {pages} {67005} (\bibinfo {year}
  {2012})}\BibitemShut {NoStop}%
\bibitem [{\citenamefont {Qiao}\ \emph {et~al.}(2017)\citenamefont {Qiao},
  \citenamefont {Li}, \citenamefont {Wang}, \citenamefont {Ruan}, \citenamefont
  {Ye}, \citenamefont {Cai}, \citenamefont {Hao}, \citenamefont {Yao},
  \citenamefont {Chen}, \citenamefont {Wu}, \citenamefont {Wang},\ and\
  \citenamefont {Liu}}]{Qiao2017}%
  \BibitemOpen
  \bibfield  {author} {\bibinfo {author} {\bibfnamefont {S.}~\bibnamefont
  {Qiao}}, \bibinfo {author} {\bibfnamefont {X.}~\bibnamefont {Li}}, \bibinfo
  {author} {\bibfnamefont {N.}~\bibnamefont {Wang}}, \bibinfo {author}
  {\bibfnamefont {W.}~\bibnamefont {Ruan}}, \bibinfo {author} {\bibfnamefont
  {C.}~\bibnamefont {Ye}}, \bibinfo {author} {\bibfnamefont {P.}~\bibnamefont
  {Cai}}, \bibinfo {author} {\bibfnamefont {Z.}~\bibnamefont {Hao}}, \bibinfo
  {author} {\bibfnamefont {H.}~\bibnamefont {Yao}}, \bibinfo {author}
  {\bibfnamefont {X.}~\bibnamefont {Chen}}, \bibinfo {author} {\bibfnamefont
  {J.}~\bibnamefont {Wu}}, \bibinfo {author} {\bibfnamefont {Y.}~\bibnamefont
  {Wang}},\ and\ \bibinfo {author} {\bibfnamefont {Z.}~\bibnamefont {Liu}},\
  }\bibfield  {title} {\bibinfo {title} {Mottness collapse in
  1\ch{T-TaS$_{2-x}$Se$_x$} transition-metal dichalcogenide: An interplay
  between localized and itinerant orbitals},\ }\href
  {https://doi.org/10.1103/PhysRevX.7.041054} {\bibfield  {journal} {\bibinfo
  {journal} {Physical Review X}\ }\textbf {\bibinfo {volume} {7}},\ \bibinfo
  {pages} {041054} (\bibinfo {year} {2017})}\BibitemShut {NoStop}%
\bibitem [{\citenamefont {Stojchevska}\ \emph {et~al.}(2014)\citenamefont
  {Stojchevska}, \citenamefont {Vaskivskyi}, \citenamefont {Mertelj},
  \citenamefont {Kusar}, \citenamefont {Svetin}, \citenamefont {Brazovskii},\
  and\ \citenamefont {Mihailovic}}]{Stojchevska2014}%
  \BibitemOpen
  \bibfield  {author} {\bibinfo {author} {\bibfnamefont {L.}~\bibnamefont
  {Stojchevska}}, \bibinfo {author} {\bibfnamefont {I.}~\bibnamefont
  {Vaskivskyi}}, \bibinfo {author} {\bibfnamefont {T.}~\bibnamefont {Mertelj}},
  \bibinfo {author} {\bibfnamefont {P.}~\bibnamefont {Kusar}}, \bibinfo
  {author} {\bibfnamefont {D.}~\bibnamefont {Svetin}}, \bibinfo {author}
  {\bibfnamefont {S.}~\bibnamefont {Brazovskii}},\ and\ \bibinfo {author}
  {\bibfnamefont {D.}~\bibnamefont {Mihailovic}},\ }\bibfield  {title}
  {\bibinfo {title} {Ultrafast switching to a stable hidden quantum state in an
  electronic crystal},\ }\href {https://doi.org/10.1126/science.1241591}
  {\bibfield  {journal} {\bibinfo  {journal} {Science}\ }\textbf {\bibinfo
  {volume} {344}},\ \bibinfo {pages} {177} (\bibinfo {year}
  {2014})}\BibitemShut {NoStop}%
\bibitem [{\citenamefont {Vaskivskyi}\ \emph {et~al.}(2015)\citenamefont
  {Vaskivskyi}, \citenamefont {Gospodaric}, \citenamefont {Brazovskii},
  \citenamefont {Svetin}, \citenamefont {Sutar}, \citenamefont {Goreshnik},
  \citenamefont {Mihailovic}, \citenamefont {Mertelj},\ and\ \citenamefont
  {Mihailovic}}]{Vaskivskyi2015}%
  \BibitemOpen
  \bibfield  {author} {\bibinfo {author} {\bibfnamefont {I.}~\bibnamefont
  {Vaskivskyi}}, \bibinfo {author} {\bibfnamefont {J.}~\bibnamefont
  {Gospodaric}}, \bibinfo {author} {\bibfnamefont {S.}~\bibnamefont
  {Brazovskii}}, \bibinfo {author} {\bibfnamefont {D.}~\bibnamefont {Svetin}},
  \bibinfo {author} {\bibfnamefont {P.}~\bibnamefont {Sutar}}, \bibinfo
  {author} {\bibfnamefont {E.}~\bibnamefont {Goreshnik}}, \bibinfo {author}
  {\bibfnamefont {I.~A.}\ \bibnamefont {Mihailovic}}, \bibinfo {author}
  {\bibfnamefont {T.}~\bibnamefont {Mertelj}},\ and\ \bibinfo {author}
  {\bibfnamefont {D.}~\bibnamefont {Mihailovic}},\ }\bibfield  {title}
  {\bibinfo {title} {Controlling the metal-to-insulator relaxation of the
  metastable hidden quantum state in 1\ch{T-TaS$_2$}},\ }\href
  {https://doi.org/10.1126/sciadv.1500168} {\bibfield  {journal} {\bibinfo
  {journal} {Science Advances}\ }\textbf {\bibinfo {volume} {1}},\ \bibinfo
  {pages} {e1500168} (\bibinfo {year} {2015})}\BibitemShut {NoStop}%
\bibitem [{\citenamefont {Gerasimenko}\ \emph {et~al.}(2019)\citenamefont
  {Gerasimenko}, \citenamefont {Karpov}, \citenamefont {Vaskivskyi},
  \citenamefont {Brazovskii},\ and\ \citenamefont
  {Mihailovic}}]{Gerasimenko2019}%
  \BibitemOpen
  \bibfield  {author} {\bibinfo {author} {\bibfnamefont {Y.~A.}\ \bibnamefont
  {Gerasimenko}}, \bibinfo {author} {\bibfnamefont {P.}~\bibnamefont {Karpov}},
  \bibinfo {author} {\bibfnamefont {I.}~\bibnamefont {Vaskivskyi}}, \bibinfo
  {author} {\bibfnamefont {S.}~\bibnamefont {Brazovskii}},\ and\ \bibinfo
  {author} {\bibfnamefont {D.}~\bibnamefont {Mihailovic}},\ }\bibfield  {title}
  {\bibinfo {title} {Intertwined chiral charge orders and topological
  stabilization of the light-induced state of a prototypical transition metal
  dichalcogenide},\ }\href {https://doi.org/10.1038/s41535-019-0172-1}
  {\bibfield  {journal} {\bibinfo  {journal} {npj Quantum Materials}\ }\textbf
  {\bibinfo {volume} {4}},\ \bibinfo {pages} {32} (\bibinfo {year}
  {2019})}\BibitemShut {NoStop}%
\bibitem [{\citenamefont {Bu}\ \emph {et~al.}(2019)\citenamefont {Bu},
  \citenamefont {Zhang}, \citenamefont {Fei}, \citenamefont {Wu}, \citenamefont
  {Zheng}, \citenamefont {Gao}, \citenamefont {Luo}, \citenamefont {Sun},\ and\
  \citenamefont {Yin}}]{Bu2019}%
  \BibitemOpen
  \bibfield  {author} {\bibinfo {author} {\bibfnamefont {K.}~\bibnamefont
  {Bu}}, \bibinfo {author} {\bibfnamefont {W.}~\bibnamefont {Zhang}}, \bibinfo
  {author} {\bibfnamefont {Y.}~\bibnamefont {Fei}}, \bibinfo {author}
  {\bibfnamefont {Z.}~\bibnamefont {Wu}}, \bibinfo {author} {\bibfnamefont
  {Y.}~\bibnamefont {Zheng}}, \bibinfo {author} {\bibfnamefont
  {J.}~\bibnamefont {Gao}}, \bibinfo {author} {\bibfnamefont {X.}~\bibnamefont
  {Luo}}, \bibinfo {author} {\bibfnamefont {Y.-P.}\ \bibnamefont {Sun}},\ and\
  \bibinfo {author} {\bibfnamefont {Y.}~\bibnamefont {Yin}},\ }\bibfield
  {title} {\bibinfo {title} {Possible strain induced mott gap collapse in
  1\ch{T-TaS2}},\ }\href {https://doi.org/10.1038/s42005-019-0247-0} {\bibfield
   {journal} {\bibinfo  {journal} {Communications Physics}\ }\textbf {\bibinfo
  {volume} {2}},\ \bibinfo {pages} {146} (\bibinfo {year} {2019})}\BibitemShut
  {NoStop}%
\bibitem [{\citenamefont {Ritschel}\ \emph {et~al.}(2013)\citenamefont
  {Ritschel}, \citenamefont {Trinckauf}, \citenamefont {Garbarino},
  \citenamefont {Hanfland}, \citenamefont {v.~Zimmermann}, \citenamefont
  {Berger}, \citenamefont {Buchner},\ and\ \citenamefont
  {Geck}}]{Ritschel2013}%
  \BibitemOpen
  \bibfield  {author} {\bibinfo {author} {\bibfnamefont {T.}~\bibnamefont
  {Ritschel}}, \bibinfo {author} {\bibfnamefont {J.}~\bibnamefont {Trinckauf}},
  \bibinfo {author} {\bibfnamefont {G.}~\bibnamefont {Garbarino}}, \bibinfo
  {author} {\bibfnamefont {M.}~\bibnamefont {Hanfland}}, \bibinfo {author}
  {\bibfnamefont {M.}~\bibnamefont {v.~Zimmermann}}, \bibinfo {author}
  {\bibfnamefont {H.}~\bibnamefont {Berger}}, \bibinfo {author} {\bibfnamefont
  {B.}~\bibnamefont {Buchner}},\ and\ \bibinfo {author} {\bibfnamefont
  {J.}~\bibnamefont {Geck}},\ }\bibfield  {title} {\bibinfo {title} {Pressure
  dependence of the charge density wave in 1\ch{T-TaS$_2$} and its relation to
  superconductivity},\ }\href {https://doi.org/10.1103/PhysRevB.87.125135}
  {\bibfield  {journal} {\bibinfo  {journal} {Physical Review B}\ }\textbf
  {\bibinfo {volume} {87}},\ \bibinfo {pages} {125135} (\bibinfo {year}
  {2013})}\BibitemShut {NoStop}%
\bibitem [{\citenamefont {Sipos}\ \emph {et~al.}(2008)\citenamefont {Sipos},
  \citenamefont {Kusmartseva}, \citenamefont {Akrap}, \citenamefont {Berger},
  \citenamefont {Forró},\ and\ \citenamefont {Tutis}}]{Sipos2008}%
  \BibitemOpen
  \bibfield  {author} {\bibinfo {author} {\bibfnamefont {B.}~\bibnamefont
  {Sipos}}, \bibinfo {author} {\bibfnamefont {A.~F.}\ \bibnamefont
  {Kusmartseva}}, \bibinfo {author} {\bibfnamefont {A.}~\bibnamefont {Akrap}},
  \bibinfo {author} {\bibfnamefont {H.}~\bibnamefont {Berger}}, \bibinfo
  {author} {\bibfnamefont {L.}~\bibnamefont {Forró}},\ and\ \bibinfo {author}
  {\bibfnamefont {E.}~\bibnamefont {Tutis}},\ }\bibfield  {title} {\bibinfo
  {title} {From mott state to superconductivity in 1\ch{T-TaS$_2$}},\ }\href
  {https://doi.org/10.1038/nmat2318} {\bibfield  {journal} {\bibinfo  {journal}
  {Nature Materials}\ }\textbf {\bibinfo {volume} {7}},\ \bibinfo {pages} {960}
  (\bibinfo {year} {2008})}\BibitemShut {NoStop}%
\bibitem [{\citenamefont {Yoshida}\ \emph {et~al.}(2014)\citenamefont
  {Yoshida}, \citenamefont {Zhang}, \citenamefont {Ye}, \citenamefont {Suzuki},
  \citenamefont {Imai}, \citenamefont {Kimura}, \citenamefont {Fujiwara},\ and\
  \citenamefont {Iwasa}}]{Yoshida2014}%
  \BibitemOpen
  \bibfield  {author} {\bibinfo {author} {\bibfnamefont {M.}~\bibnamefont
  {Yoshida}}, \bibinfo {author} {\bibfnamefont {Y.}~\bibnamefont {Zhang}},
  \bibinfo {author} {\bibfnamefont {J.}~\bibnamefont {Ye}}, \bibinfo {author}
  {\bibfnamefont {R.}~\bibnamefont {Suzuki}}, \bibinfo {author} {\bibfnamefont
  {Y.}~\bibnamefont {Imai}}, \bibinfo {author} {\bibfnamefont {S.}~\bibnamefont
  {Kimura}}, \bibinfo {author} {\bibfnamefont {A.}~\bibnamefont {Fujiwara}},\
  and\ \bibinfo {author} {\bibfnamefont {Y.}~\bibnamefont {Iwasa}},\ }\bibfield
   {title} {\bibinfo {title} {Controlling charge-density-wave states in
  nano-thick crystals of 1\ch{T-TaS$_2$}},\ }\href
  {https://doi.org/10.1038/srep07302} {\bibfield  {journal} {\bibinfo
  {journal} {Scientific Reports}\ }\textbf {\bibinfo {volume} {4}},\ \bibinfo
  {pages} {7302} (\bibinfo {year} {2014})}\BibitemShut {NoStop}%
\bibitem [{\citenamefont {Yoshida}\ \emph {et~al.}(2015)\citenamefont
  {Yoshida}, \citenamefont {Suzuki}, \citenamefont {Zhang}, \citenamefont
  {Nakano},\ and\ \citenamefont {Iwasa}}]{Yoshida2015}%
  \BibitemOpen
  \bibfield  {author} {\bibinfo {author} {\bibfnamefont {M.}~\bibnamefont
  {Yoshida}}, \bibinfo {author} {\bibfnamefont {R.}~\bibnamefont {Suzuki}},
  \bibinfo {author} {\bibfnamefont {Y.}~\bibnamefont {Zhang}}, \bibinfo
  {author} {\bibfnamefont {M.}~\bibnamefont {Nakano}},\ and\ \bibinfo {author}
  {\bibfnamefont {Y.}~\bibnamefont {Iwasa}},\ }\bibfield  {title} {\bibinfo
  {title} {Memristive phase switching in two-dimensional 1\ch{T-TaS$_2$}
  crystals},\ }\href {https://doi.org/10.1126/sciadv.1500606} {\bibfield
  {journal} {\bibinfo  {journal} {Science Advances}\ }\textbf {\bibinfo
  {volume} {1}},\ \bibinfo {pages} {e1500606} (\bibinfo {year}
  {2015})}\BibitemShut {NoStop}%
\bibitem [{\citenamefont {Ishiguro}\ and\ \citenamefont
  {Sato}(1991)}]{Ishiguro1991}%
  \BibitemOpen
  \bibfield  {author} {\bibinfo {author} {\bibfnamefont {T.}~\bibnamefont
  {Ishiguro}}\ and\ \bibinfo {author} {\bibfnamefont {H.}~\bibnamefont
  {Sato}},\ }\bibfield  {title} {\bibinfo {title} {Electron microscopy of phase
  transformations in 1\ch{T-TaS2}},\ }\href
  {https://doi.org/10.1103/PhysRevB.44.2046} {\bibfield  {journal} {\bibinfo
  {journal} {Physical Review B}\ }\textbf {\bibinfo {volume} {44}},\ \bibinfo
  {pages} {2046} (\bibinfo {year} {1991})}\BibitemShut {NoStop}%
\bibitem [{\citenamefont {Momma}\ and\ \citenamefont
  {Izumi}(2011)}]{Momma2011}%
  \BibitemOpen
  \bibfield  {author} {\bibinfo {author} {\bibfnamefont {K.}~\bibnamefont
  {Momma}}\ and\ \bibinfo {author} {\bibfnamefont {F.}~\bibnamefont {Izumi}},\
  }\bibfield  {title} {\bibinfo {title} {Vesta.3 for three-dimensional
  visualization of crystal, volumetric and morphology data},\ }\href
  {https://doi.org/https://doi.org/10.1107/S0021889811038970} {\bibfield
  {journal} {\bibinfo  {journal} {Journal of Applied Crystallography}\ }\textbf
  {\bibinfo {volume} {44}},\ \bibinfo {pages} {1272} (\bibinfo {year}
  {2011})}\BibitemShut {NoStop}%
\bibitem [{\citenamefont {Scruby}\ \emph {et~al.}(1975)\citenamefont {Scruby},
  \citenamefont {Williams},\ and\ \citenamefont {Parry}}]{Scruby1975}%
  \BibitemOpen
  \bibfield  {author} {\bibinfo {author} {\bibfnamefont {C.~B.}\ \bibnamefont
  {Scruby}}, \bibinfo {author} {\bibfnamefont {P.~M.}\ \bibnamefont
  {Williams}},\ and\ \bibinfo {author} {\bibfnamefont {G.~S.}\ \bibnamefont
  {Parry}},\ }\bibfield  {title} {\bibinfo {title} {The role of charge density
  waves in structural transformations of 1\ch{T} \ch{TaS$_2$}},\ }\href
  {https://doi.org/10.1080/14786437508228930} {\bibfield  {journal} {\bibinfo
  {journal} {The Philosophical Magazine: A Journal of Theoretical Experimental
  and Applied Physics}\ }\textbf {\bibinfo {volume} {31}},\ \bibinfo {pages}
  {255} (\bibinfo {year} {1975})}\BibitemShut {NoStop}%
\bibitem [{\citenamefont {Shao}\ \emph {et~al.}(2016)\citenamefont {Shao},
  \citenamefont {Xiao}, \citenamefont {Lu}, \citenamefont {Lv}, \citenamefont
  {Li}, \citenamefont {Zhu},\ and\ \citenamefont {Sun}}]{Shao2016}%
  \BibitemOpen
  \bibfield  {author} {\bibinfo {author} {\bibfnamefont {D.~F.}\ \bibnamefont
  {Shao}}, \bibinfo {author} {\bibfnamefont {R.~C.}\ \bibnamefont {Xiao}},
  \bibinfo {author} {\bibfnamefont {W.~J.}\ \bibnamefont {Lu}}, \bibinfo
  {author} {\bibfnamefont {H.~Y.}\ \bibnamefont {Lv}}, \bibinfo {author}
  {\bibfnamefont {J.~Y.}\ \bibnamefont {Li}}, \bibinfo {author} {\bibfnamefont
  {X.~B.}\ \bibnamefont {Zhu}},\ and\ \bibinfo {author} {\bibfnamefont {Y.~P.}\
  \bibnamefont {Sun}},\ }\bibfield  {title} {\bibinfo {title} {Manipulating
  charge density waves in 1\ch{T-TaS$_2$} by charge-carrier doping: A
  first-principles investigation},\ }\href
  {https://doi.org/10.1103/PhysRevB.94.125126} {\bibfield  {journal} {\bibinfo
  {journal} {Physical Review B}\ }\textbf {\bibinfo {volume} {94}},\ \bibinfo
  {pages} {125126} (\bibinfo {year} {2016})}\BibitemShut {NoStop}%
\bibitem [{\citenamefont {Wang}\ \emph {et~al.}(2019)\citenamefont {Wang},
  \citenamefont {Dietzel},\ and\ \citenamefont {Schirmeisen}}]{Wang2019}%
  \BibitemOpen
  \bibfield  {author} {\bibinfo {author} {\bibfnamefont {W.}~\bibnamefont
  {Wang}}, \bibinfo {author} {\bibfnamefont {D.}~\bibnamefont {Dietzel}},\ and\
  \bibinfo {author} {\bibfnamefont {A.}~\bibnamefont {Schirmeisen}},\
  }\bibfield  {title} {\bibinfo {title} {Lattice discontinuities of
  1\ch{T-TaS$_2$} across first order charge density wave phase transitions},\
  }\href {https://doi.org/10.1038/s41598-019-43307-2} {\bibfield  {journal}
  {\bibinfo  {journal} {Scientific Reports}\ }\textbf {\bibinfo {volume} {9}},\
  \bibinfo {pages} {7066} (\bibinfo {year} {2019})}\BibitemShut {NoStop}%
\bibitem [{\citenamefont {Tsen}\ \emph {et~al.}(2015)\citenamefont {Tsen},
  \citenamefont {Hovden}, \citenamefont {Wang}, \citenamefont {Kim},
  \citenamefont {Okamoto}, \citenamefont {Spoth}, \citenamefont {Liu},
  \citenamefont {Lu}, \citenamefont {Sun}, \citenamefont {Hone}, \citenamefont
  {Kourkoutis}, \citenamefont {Kim},\ and\ \citenamefont
  {Pasupathy}}]{Tsen2015}%
  \BibitemOpen
  \bibfield  {author} {\bibinfo {author} {\bibfnamefont {A.~W.}\ \bibnamefont
  {Tsen}}, \bibinfo {author} {\bibfnamefont {R.}~\bibnamefont {Hovden}},
  \bibinfo {author} {\bibfnamefont {D.}~\bibnamefont {Wang}}, \bibinfo {author}
  {\bibfnamefont {Y.~D.}\ \bibnamefont {Kim}}, \bibinfo {author} {\bibfnamefont
  {J.}~\bibnamefont {Okamoto}}, \bibinfo {author} {\bibfnamefont {K.~A.}\
  \bibnamefont {Spoth}}, \bibinfo {author} {\bibfnamefont {Y.}~\bibnamefont
  {Liu}}, \bibinfo {author} {\bibfnamefont {W.}~\bibnamefont {Lu}}, \bibinfo
  {author} {\bibfnamefont {Y.}~\bibnamefont {Sun}}, \bibinfo {author}
  {\bibfnamefont {J.~C.}\ \bibnamefont {Hone}}, \bibinfo {author}
  {\bibfnamefont {L.~F.}\ \bibnamefont {Kourkoutis}}, \bibinfo {author}
  {\bibfnamefont {P.}~\bibnamefont {Kim}},\ and\ \bibinfo {author}
  {\bibfnamefont {A.~N.}\ \bibnamefont {Pasupathy}},\ }\bibfield  {title}
  {\bibinfo {title} {Structure and control of charge density waves in
  two-dimensional 1\ch{T-TaS$_2$}},\ }\href
  {https://doi.org/10.1073/pnas.1512092112} {\bibfield  {journal} {\bibinfo
  {journal} {Proceedings of the National Academy of Sciences}\ }\textbf
  {\bibinfo {volume} {112}},\ \bibinfo {pages} {15054} (\bibinfo {year}
  {2015})}\BibitemShut {NoStop}%
\bibitem [{\citenamefont {Ishioka}\ \emph {et~al.}(2010)\citenamefont
  {Ishioka}, \citenamefont {Liu}, \citenamefont {Shimatake}, \citenamefont
  {Kurosawa}, \citenamefont {Ichimura}, \citenamefont {Toda}, \citenamefont
  {Oda},\ and\ \citenamefont {Tanda}}]{Ishioka2010}%
  \BibitemOpen
  \bibfield  {author} {\bibinfo {author} {\bibfnamefont {J.}~\bibnamefont
  {Ishioka}}, \bibinfo {author} {\bibfnamefont {Y.~H.}\ \bibnamefont {Liu}},
  \bibinfo {author} {\bibfnamefont {K.}~\bibnamefont {Shimatake}}, \bibinfo
  {author} {\bibfnamefont {T.}~\bibnamefont {Kurosawa}}, \bibinfo {author}
  {\bibfnamefont {K.}~\bibnamefont {Ichimura}}, \bibinfo {author}
  {\bibfnamefont {Y.}~\bibnamefont {Toda}}, \bibinfo {author} {\bibfnamefont
  {M.}~\bibnamefont {Oda}},\ and\ \bibinfo {author} {\bibfnamefont
  {S.}~\bibnamefont {Tanda}},\ }\bibfield  {title} {\bibinfo {title} {Chiral
  charge-density waves},\ }\href
  {https://doi.org/10.1103/PhysRevLett.105.176401} {\bibfield  {journal}
  {\bibinfo  {journal} {Physical Review Letters}\ }\textbf {\bibinfo {volume}
  {105}},\ \bibinfo {pages} {176401} (\bibinfo {year} {2010})}\BibitemShut
  {NoStop}%
\bibitem [{\citenamefont {Gao}\ \emph {et~al.}(2021)\citenamefont {Gao},
  \citenamefont {Zhang}, \citenamefont {Si}, \citenamefont {Luo}, \citenamefont
  {Yan}, \citenamefont {Jiang}, \citenamefont {Wang}, \citenamefont {Lv},
  \citenamefont {Tong}, \citenamefont {Song}, \citenamefont {Zhu},
  \citenamefont {Lu}, \citenamefont {Yin},\ and\ \citenamefont
  {Sun}}]{Gao2021}%
  \BibitemOpen
  \bibfield  {author} {\bibinfo {author} {\bibfnamefont {J.~J.}\ \bibnamefont
  {Gao}}, \bibinfo {author} {\bibfnamefont {W.~H.}\ \bibnamefont {Zhang}},
  \bibinfo {author} {\bibfnamefont {J.~G.}\ \bibnamefont {Si}}, \bibinfo
  {author} {\bibfnamefont {X.}~\bibnamefont {Luo}}, \bibinfo {author}
  {\bibfnamefont {J.}~\bibnamefont {Yan}}, \bibinfo {author} {\bibfnamefont
  {Z.~Z.}\ \bibnamefont {Jiang}}, \bibinfo {author} {\bibfnamefont
  {W.}~\bibnamefont {Wang}}, \bibinfo {author} {\bibfnamefont {H.~Y.}\
  \bibnamefont {Lv}}, \bibinfo {author} {\bibfnamefont {P.}~\bibnamefont
  {Tong}}, \bibinfo {author} {\bibfnamefont {W.~H.}\ \bibnamefont {Song}},
  \bibinfo {author} {\bibfnamefont {X.~B.}\ \bibnamefont {Zhu}}, \bibinfo
  {author} {\bibfnamefont {W.~J.}\ \bibnamefont {Lu}}, \bibinfo {author}
  {\bibfnamefont {Y.}~\bibnamefont {Yin}},\ and\ \bibinfo {author}
  {\bibfnamefont {Y.~P.}\ \bibnamefont {Sun}},\ }\bibfield  {title} {\bibinfo
  {title} {Chiral charge density waves induced by \ch{Ti}-doping in
  1\ch{T-TaS2}},\ }\href {https://doi.org/10.1063/5.0052240} {\bibfield
  {journal} {\bibinfo  {journal} {Applied Physics Letters}\ }\textbf {\bibinfo
  {volume} {118}},\ \bibinfo {pages} {213105} (\bibinfo {year}
  {2021})}\BibitemShut {NoStop}%
\bibitem [{\citenamefont {Klosinski}\ \emph {et~al.}(2021)\citenamefont
  {Klosinski}, \citenamefont {Oles}, \citenamefont {Efthimia~Agrapidis},
  \citenamefont {van Wezel},\ and\ \citenamefont {Wohlfeld}}]{Klosinski2021}%
  \BibitemOpen
  \bibfield  {author} {\bibinfo {author} {\bibfnamefont {A.}~\bibnamefont
  {Klosinski}}, \bibinfo {author} {\bibfnamefont {A.~M.}\ \bibnamefont {Oles}},
  \bibinfo {author} {\bibfnamefont {C.}~\bibnamefont {Efthimia~Agrapidis}},
  \bibinfo {author} {\bibfnamefont {J.}~\bibnamefont {van Wezel}},\ and\
  \bibinfo {author} {\bibfnamefont {K.}~\bibnamefont {Wohlfeld}},\ }\bibfield
  {title} {\bibinfo {title} {Chalcogenic orbital density waves in the weak- and
  strong-coupling limit},\ }\href {https://doi.org/10.1103/PhysRevB.103.235123}
  {\bibfield  {journal} {\bibinfo  {journal} {Physical Review B}\ }\textbf
  {\bibinfo {volume} {103}},\ \bibinfo {pages} {235123} (\bibinfo {year}
  {2021})}\BibitemShut {NoStop}%
\bibitem [{\citenamefont {van Wezel}(2011)}]{vanWezel2011}%
  \BibitemOpen
  \bibfield  {author} {\bibinfo {author} {\bibfnamefont {J.}~\bibnamefont {van
  Wezel}},\ }\bibfield  {title} {\bibinfo {title} {Chirality and orbital order
  in charge density waves},\ }\href
  {https://doi.org/10.1209/0295-5075/96/67011} {\bibfield  {journal} {\bibinfo
  {journal} {EPL (Europhysics Letters)}\ }\textbf {\bibinfo {volume} {96}},\
  \bibinfo {pages} {67011} (\bibinfo {year} {2011})}\BibitemShut {NoStop}%
\bibitem [{\citenamefont {Peng}\ \emph {et~al.}(2021)\citenamefont {Peng},
  \citenamefont {Guo}, \citenamefont {Xiao}, \citenamefont {Li}, \citenamefont
  {Strempfer}, \citenamefont {Choi}, \citenamefont {Yan}, \citenamefont {Luo},
  \citenamefont {Huang}, \citenamefont {Jia}, \citenamefont {Janson},
  \citenamefont {Abbamonte}, \citenamefont {van~den Brink},\ and\ \citenamefont
  {van Wezel}}]{Peng2021}%
  \BibitemOpen
  \bibfield  {author} {\bibinfo {author} {\bibfnamefont {Y.}~\bibnamefont
  {Peng}}, \bibinfo {author} {\bibfnamefont {X.}~\bibnamefont {Guo}}, \bibinfo
  {author} {\bibfnamefont {Q.}~\bibnamefont {Xiao}}, \bibinfo {author}
  {\bibfnamefont {Q.}~\bibnamefont {Li}}, \bibinfo {author} {\bibfnamefont
  {J.}~\bibnamefont {Strempfer}}, \bibinfo {author} {\bibfnamefont
  {Y.}~\bibnamefont {Choi}}, \bibinfo {author} {\bibfnamefont {D.}~\bibnamefont
  {Yan}}, \bibinfo {author} {\bibfnamefont {H.}~\bibnamefont {Luo}}, \bibinfo
  {author} {\bibfnamefont {Y.}~\bibnamefont {Huang}}, \bibinfo {author}
  {\bibfnamefont {S.}~\bibnamefont {Jia}}, \bibinfo {author} {\bibfnamefont
  {O.}~\bibnamefont {Janson}}, \bibinfo {author} {\bibfnamefont
  {P.}~\bibnamefont {Abbamonte}}, \bibinfo {author} {\bibfnamefont
  {M.}~\bibnamefont {van~den Brink}},\ and\ \bibinfo {author} {\bibfnamefont
  {J.}~\bibnamefont {van Wezel}},\ }\bibfield  {title} {\bibinfo {title}
  {Observation of orbital order in the van der waals material
  1\ch{T-TiSe$_2$}},\ }\href@noop {} {\bibfield  {journal} {\bibinfo  {journal}
  {arXiv:2105.13195}\ } (\bibinfo {year} {2021})}\BibitemShut {NoStop}%
\bibitem [{\citenamefont {Kratochvilova}\ \emph {et~al.}(2017)\citenamefont
  {Kratochvilova}, \citenamefont {Hillier}, \citenamefont {Wildes},
  \citenamefont {Wang}, \citenamefont {Cheong},\ and\ \citenamefont
  {Park}}]{Kratochvilova2017}%
  \BibitemOpen
  \bibfield  {author} {\bibinfo {author} {\bibfnamefont {M.}~\bibnamefont
  {Kratochvilova}}, \bibinfo {author} {\bibfnamefont {A.~D.}\ \bibnamefont
  {Hillier}}, \bibinfo {author} {\bibfnamefont {A.~R.}\ \bibnamefont {Wildes}},
  \bibinfo {author} {\bibfnamefont {L.}~\bibnamefont {Wang}}, \bibinfo {author}
  {\bibfnamefont {S.-W.}\ \bibnamefont {Cheong}},\ and\ \bibinfo {author}
  {\bibfnamefont {J.-G.}\ \bibnamefont {Park}},\ }\bibfield  {title} {\bibinfo
  {title} {The low-temperature highly correlated quantum phase in the
  charge-density-wave 1\ch{T-TaS2} compound},\ }\href
  {https://doi.org/10.1038/s41535-017-0048-1} {\bibfield  {journal} {\bibinfo
  {journal} {npj Quantum Materials}\ }\textbf {\bibinfo {volume} {2}},\
  \bibinfo {pages} {42} (\bibinfo {year} {2017})}\BibitemShut {NoStop}%
\end{thebibliography}%

\end{document}


\section{Supplemental Material}
We examine the simulated topography shown in Figure 3a of the manuscript. Figure S1a shows the simulated average CDW lattice (only the average CDW signal is included – signals associated with peaks encircled in white in Figure 1c). This is a purely hexagonal lattice. Figure S1b shows the simulated true CDW lattice (in addition to the average CDW signal, signals associated with the two brightest satellite peaks around each CDW peak of the FFT are included). The domain structure of the NC-CDW state is clearly seen.

Figure S1c shows arrows pointing from the average CDW lattice to the true CDW lattice. These represent the shift of the true CDW maximum in the topography relative to the average CDW maximum location. The length of the arrows are proportional to the size of the shift. Overlaid in blue are circles as a reference to the eye showing regions of clear chiral shifts. Overlaying the same circles in the topography (Figure S1d) shows these circles align well with the domains of the NC-CDW state and indicate the ability to use the shift arrows to identify domain locations. Domain edges occur where neighboring arrows oppose one another.  Quantifying the true CDW shift from the average CDW maximum location, we find an average displacement of 0.82 {\AA} with a maximum displacement of 1.35 {\AA}. These are close to the average displacement of 0.72 {\AA} with a maximum displacement of 1.70 {\AA} found when analyzing the STM-acquired data.

Analysis of the simulated topography shows clearly the intradomain chiral nature of the NC-CDW state of 1T-TaS$_2$. As seen in the analysis of the STM-acquired data, there is a counter-clockwise vortex-like shift of the true CDW maxima relative to the average CDW locations.

\begin{figure}[ht]
\includegraphics[clip=true,width=0.6\columnwidth]{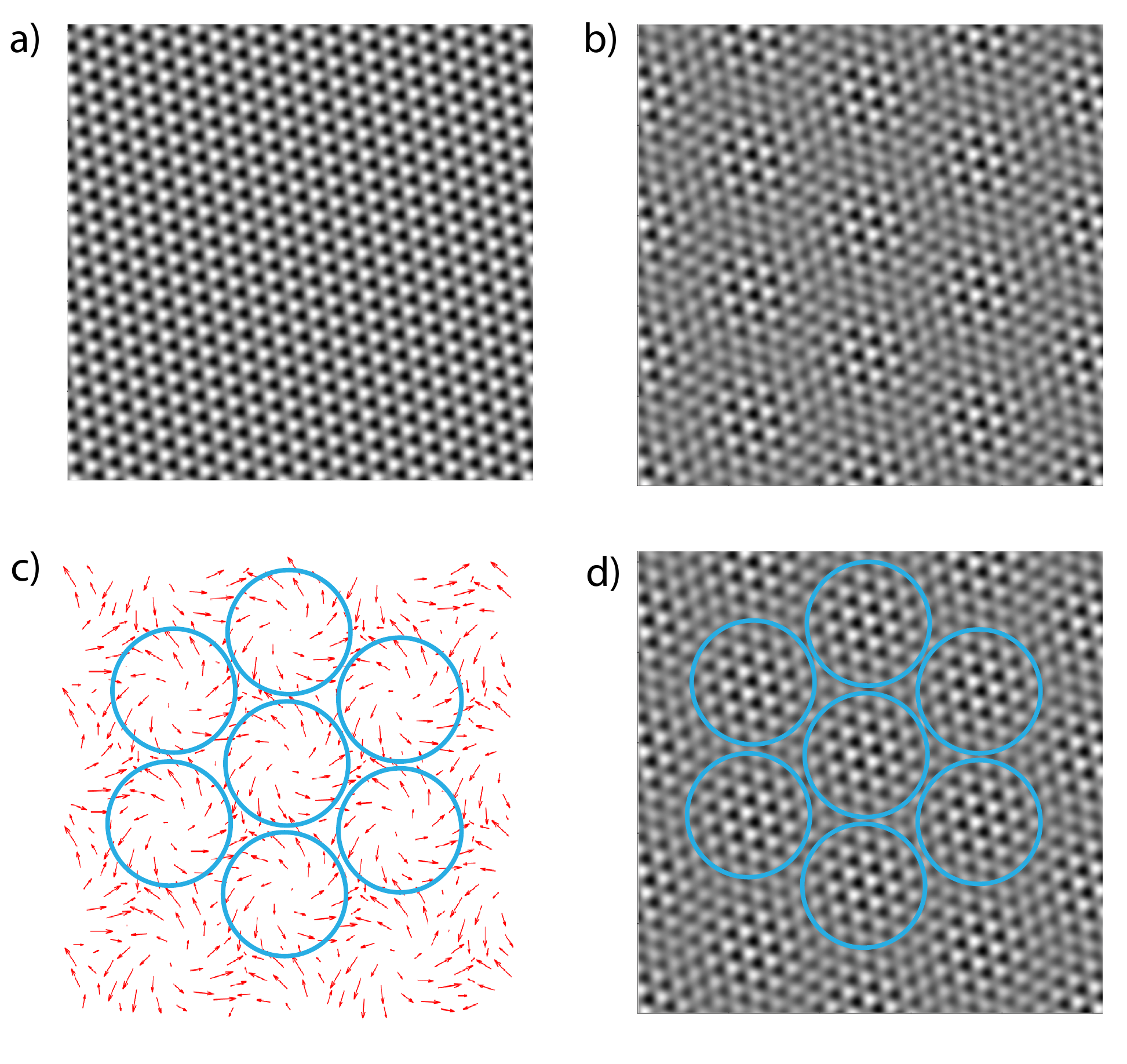}
\caption{a) Average CDW lattice: Simulated topography from Figure 3a which includes only the average CDW signal (no satellite peak signals). b) True CDW lattice: Simulated topography including average CDW signal and the two most intense satellite peaks around each CDW peak in the FFT shown in Figure 1c. c) Shift arrows: The arrows start at the average CDW location (Figure S1a) and point to the true CDW location (Figure S1b). Blue circles are included as a guide to the eye to emphasize the chiral shifts. d) Blue circles overlaid on the true CDW lattice illustrating the chiral shifts seen in c) are associated with the domains of the NC-CDW state.}
\label{fig:FigS1}
\end{figure}